\documentclass[prl, twocolumn, showpacs, letter]{revtex4}
\usepackage{pslatex}
\usepackage{amsmath}
\usepackage{amssymb}
\usepackage{amsfonts}
\usepackage{amsthm}
\usepackage{graphicx}

\newcommand{\Bracket}[1]{\left\langle #1 \right\rangle}

\begin{document}
\bibliographystyle{apsrev}

\title{Anomalous Diffusion on Random Graphs}



\author{Joseph Snider}
\email[]{jsnider@uci.edu}
\affiliation{Department of Physics and Astronomy, University of
  California Irvine, Irvine, CA 92697-4575}

\author{Clare C. Yu}
\email[]{cyu@uci.edu}
\affiliation{Department of Physics and Astronomy, University of
  California Irvine, Irvine,
  CA 92697-4575}

\date{\today}

\begin{abstract}
We show that anomalous diffusion can result when the steps of a
random walk are not statistically independent. We present an
algorithm that counts all the possible paths of particles
diffusing on random graphs with arbitrary degree distribution.
Using this to calculate the mean square displacement, we show that
in sharp contrast to continua, random walks on random graphs can
exhibit anomalous behavior and yet have well-defined and
predictable properties.
\end{abstract}


\maketitle

\unitlength 1mm


It is well known that a random walk on a continuous medium leads
to a mean square displacement $\Bracket{x^{2}}$ that is linear in
time as long as there are no infinitely large steps and all steps
are statistically independent. L$\acute{e}$vy showed that
anomalous diffusion results when the former assumption is
violated. In this paper we show that anomalous diffusion results
when the latter assumption is violated but the former remains, by
considering random walks on random graphs.

Random graphs have been very successfully used to describe many
diverse systems.  For example, the spreading of diseases through a
population has been modelled where people are the vertices and
contact between them the edges \cite{Newman2000,Liljeros1998}. The
internet can be thought of as a graph with web sites as vertices
and links as edges, leading to estimates of the number of clicks
to surf between any two random sites and other quantities of
interest \cite{Lawrence1998,Albert2000}. In general, any system
with interacting parts, which encompasses a vast array of diverse
systems, can be mapped onto a graph \cite{Albert2002}.

Along with to statics, dynamic properties of random graphs are of
interest for many cases like protein folding \cite{Sokolov1997}
and glassy relaxation \cite{Bray1988}. In this paper we consider
particles diffusing along the edges of a random graph and develop
a method for calculating diffusion on a random graph using an
algorithm that counts all the paths.


\begin{figure}[tbp]
   \begin{center}
      \includegraphics[width=2in]{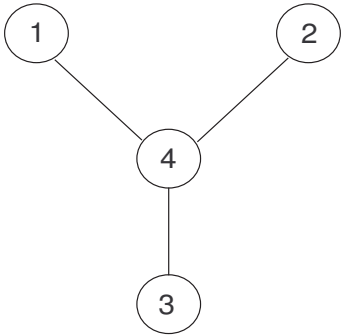}
      \caption{A simple graph. The circles represent vertices and the
      lines represent edges.}
      \label{fig:Graph3Star}
   \end{center}
\end{figure}

The following is a brief introduction to graph theory.  For a more
comprehensive overview see, for example, Godsil and Royle
\cite{Godsil2001}. Graphs consist of two sets: vertices, which are
points, and edges, which join two vertices. In more mathematical
terms, define for some index set $\mathbb{I}$, the vertex set

\begin{equation}
   V = \{v_{i} \mid i \in \mathbb{I} \}.
\end{equation}

Then, define the edge set as a subset of the pairs of all vertices

\begin{equation}
   E = \{ (v_{j}, v_{k}) \mid v_{j}, v_{k} \in V \} \subseteq V \times
   V.
\end{equation}

These two sets can be thought of as points and lines in a plane as
in Figure \ref{fig:Graph3Star}. The pairs of vertices defining an
edge are called endpoints of the edge. There are no real
restrictions on the sets of vertices and edges, making graphs very
general constructions.

Like any field, graph theory has its own lingo so here are some
definitions. The degree of a vertex is the number of edges with at
least one endpoint at the vertex.  The degree distribution $p(k)$
is the probability that a randomly selected vertex will have
degree $k$.  Two vertices are adjacent if there is an edge joining
them, i.e., in Figure \ref{fig:Graph3Star}, vertices $1$ and $4$
are adjacent but $1$ and $2$ are not. A path or walk on a graph is
a set of edges connecting two vertices. For example, in Figure
\ref{fig:Graph3Star} paths from $1$ to $2$ would be the the set of
edges $\{(1,4), (4,2)\}$ or $\{(1,4), (4,3), (3,4), (4,2)\}$.

Now, we have enough definitions to consider why it is possible for
a random graph to violate the assumption that all steps of a walk
are independent. Consider the graph in Figure
\ref{fig:Graph3Star}.  There are three vertices with one edge and
one vertex with three edges. Thus, the degree distribution is
\begin{equation}
   p(k) =
   \begin{cases}
      \frac{3}{4}&  k=1\\
      \frac{1}{4}&  k=3\\
      0          &  \text{otherwise}.
   \end{cases}
\end{equation}

Next, put a walker on that graph that can move from vertex to
vertex along the available edges.  For a walker's first step there
is a probability $p(k)$ that it will have $k$ choices. However,
for the second step this is not necessarily the case. Still
looking at the graph in Figure \ref{fig:Graph3Star}, a walker
starting at vertex $1$, $2$, or $3$ will end up at vertex $4$
after one step where it will have three choices of edges on which
to leave. A walker starting at vertex $4$ can go to $1$, $2$, or
$3$ where it will have one choice of edges on which to leave.
Thus, there are a total of six nearest neighbors, one each from
vertices $1$, $2$, and $3$ and three from vertex $4$, and the
degree distribution for them, $q(k)$, is

\begin{equation}
   q(k) =
   \begin{cases}
      \frac{3}{6} = \frac{1}{2}&  k=1\\
      \frac{3}{6} = \frac{1}{2}&  k=3\\
      0          &  \text{otherwise}.
   \end{cases}
\end{equation}

Thus, since $p(k) \neq q(k)$ the degree distribution is a function
of the number of steps taken.  This means that the number of
choices available to a walker on a graph will depend on how far it
has gone. Therefore, the assumption that all steps are
uncorrelated is violated in this graph, and it can be violated
generally in graphs. Note that it is not necessarily violated; for
example, in a graph where all vertices have the same degree,
called a regular graph, $p(k)$ and $q(k)$ are the same.


Moving on, some more tools for dealing with graphs will be
introduced. First, define $d(\mu,\nu)$ to be the distance between
two vertices $\mu$ and $\nu$ which is the size of the smallest set
of edges making a path starting at $\mu$ and ending at $\nu$. For
example in Figure \ref{fig:Graph3Star}, $d(1,2)=2$ and $d(1,4)=1$.
Unless otherwise noted, all distances mentioned henceforth refer
to this graph distance.

Next, define the adjacency matrix $\mathbb{A}$ of the graph
$G=(V,E)$ to be a square $|V| \times |V|$ matrix with
$\mathbb{A}_{\mu \nu}$ equal to the number of edges joining the
vertices $\mu$ and $\nu$. Note that so long as there are no edges
that connect a vertex to itself, i.e. loops, the diagonal elements
of $\mathbb{A}$ must be zero, and an edge connecting vertex $\mu$
to $\nu$ also connects $\nu$ to $\mu$ so $\mathbb{A}$ is
symmetric.

As an example, consider the graph in Figure \ref{fig:Graph3Star}
which has adjacency matrix

\begin{equation}
   \mathbb{A} =
      \begin{pmatrix}
         0 & 0 & 0 & 1 \\
         0 & 0 & 0 & 1 \\
         0 & 0 & 0 & 1 \\
         1 & 1 & 1 & 0
      \end{pmatrix}.
\end{equation}

The adjacency matrix of a graph can be used to count all paths on
the graph of a given length.  By definition the element
$\mathbb{A}_{\mu \nu}$ gives the number of paths of length one
from vertex $\mu$ to $\nu$.  It can be shown by induction that
$\left(\mathbb{A}^{n}\right)_{\mu \nu}$ gives the exact number of
paths of length n from $\mu$ to $\nu$ \cite{Godsil2001}. For
example, looking at the graph in Figure \ref{fig:Graph3Star}, the
number of paths of length exactly two between any two vertices are
given by the elements of

\begin{equation}
   \mathbb{A}^{2} =
      \begin{pmatrix}
         1 & 1 & 1 & 0 \\
         1 & 1 & 1 & 0 \\
         1 & 1 & 1 & 0 \\
         0 & 0 & 0 & 3
      \end{pmatrix}.
\end{equation}
The reader is encouraged to verify this by inspection. Note that
edges can be traversed more than once and are multidirectional.


\begin{figure}[tbp]
   \begin{center}
      \includegraphics[width=2in]{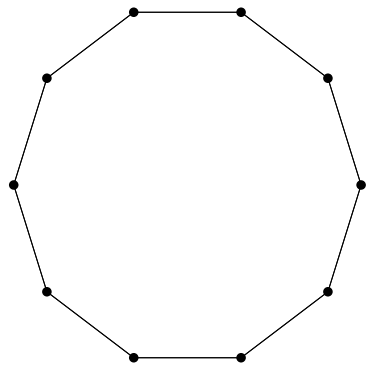}
\caption{A regular ring graph.  As the size grows, a walk on this
graph approaches one dimensional diffusion, i.e. the drunkard's
walk.}
      \label{fig:GraphRegular2}
   \end{center}
\end{figure}

Consider the regular graph in Figure \ref{fig:GraphRegular2}. A
random walk on this graph is just the well-known drunkard's walk
with periodic boundary conditions which is not anomalous.  Let
$\Bracket{x^{2}(n)}$ denote the average distance squared as a
function of $n$, where $n$ is the number of steps taken and $x$ is
the distance measured from some starting vertex on the graph. One
way to calculate $\Bracket{x^{2}(n)}$ is to choose some starting
vertex and count the fraction of paths of length $n$ that are only
a distance $m<n$ away from the start, and then average over all
possible starting vertices. Call this quantity $p_{m}^{n}$.
Looking at the drunkard's walk graph, Figure
\ref{fig:GraphRegular2}, choose any vertex as the start since they
are all equivalent. Then, there are two paths of length $1$, one
step clockwise and one step counterclockwise, both of which are
distance $1$ away from the start, so $p_{m}^{1} = \delta _{m1}$.
There are four paths of length $2$ of which two are one step out
and one step back, giving a distance of $0$, and two are two steps
out clockwise or counterclockwise, giving distance $2$. So
$p_{m}^{2}$ is given by

\begin{equation}
   p_{m}^{2} =
   \begin{cases}
      \frac{1}{2} &  m=0\\
      \frac{1}{2} &  m=2\\
      0          &  \text{otherwise}.
   \end{cases}
\end{equation}

In general, as the number of steps increases $p_{m}^{n}$ will
converge to a binomial distribution leading to the standard result
\begin{equation}
\begin{split}\label{eq:x_squared}
   \Bracket{x^{2}(n)} &= \sum_{m}m^{2}p_{m}^{n} \\
                      &= Dn,
\end{split}
\end{equation}
where $D$ is the diffusion coefficient \cite{Reif1965}.

The central limit theorem says that changing the step size of the
above walk only changes the diffusion coefficient, not the linear
dependence on $n$.  Thus, if the dependence on $n$ is the only
quantity of interest, graphs can be used to calculate it even
though they ignore all spatial distances and only concentrate on
the number of choices available at each step.

Using an approach similar to the above example, the adjacency
matrix can be used to calculate $p_{m}^{n}$, and hence the
dependence of $\Bracket{x^{2}}$ on $n$ for any general graph.
Since the adjacency matrix $\mathbb{A}$ can be used to calculate
the exact number of paths of a given length between any two
vertices, we can enumerate all paths on the graph and find
$p_{m}^{n}$ using the following algorithm:
\begin{enumerate}
   \item Let $n$ be the current path length, starting with $1$.
   \item Raise $\mathbb{A}$ to the $n^{th}$ power.
   \item Find the number of paths between any two vertices by
   looking at the elements of $\mathbb{A}^{n}$.
   \item Keep track of the minimum distance between any two vertices by
   noting if this is the smallest $n$ such that a path exists.
   \item For every $m \leq n$ find $p_{m}^{n}$.
   \item Increment $n$ and repeat.
\end{enumerate}
Once we have the distribution $p_{m}^{n}$, all the random walk
properties can be calculated. In particular, equation
\ref{eq:x_squared} allows the calculation of $\Bracket{x^{2}(n)}$.


\begin{figure}[tbp]
   \begin{center}
      \includegraphics[width=3in]{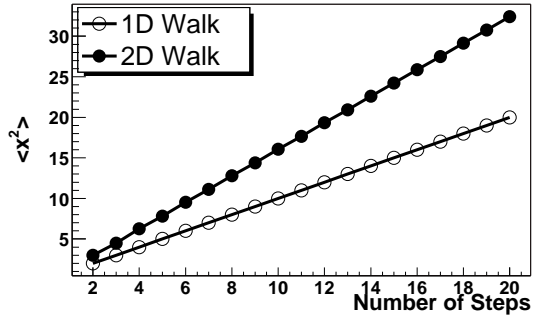}
\caption{$\Bracket{x^{2}(n)}$ as calculated with adjacency
matrices for walks on a one dimensional ring with 1000 vertices
and for a two dimensional square lattice of size $40$x$40$ with
periodic boundary conditions. Note that $\Bracket{x^{2}(n)}$ is
linear in both cases as expected, but the slope changes with the
details of the system. The lines are guides to the eye.}
      \label{fig:PlotDrunkenWalk}
   \end{center}
\end{figure}

The method presented here for enumerating paths to calculate
random walk properties reduces to just a problem of matrix
multiplication. Thus, it lends itself very well to calculation by
computer where efficient and easy to use matrix multiplication
algorithms are readily available.

As a first test of this method, diffusion on a ring should be
equivalent to the drunkard's walk for large enough rings. The same
should be true for any graph representing a regular lattice.
Figure \ref{fig:PlotDrunkenWalk} shows a calculation of
$\Bracket{x^{2}(n)}$ using adjacency matrices as described above
for a ring graph with $1000$ vertices and a periodic two
dimensional square lattice of size 40 by 40. From the plot it is
apparent that this adjacency matrix algorithm gives the correct
exponent of $1$, i.e. $\Bracket{x^{2}(n)}\sim n$.  The slope of
the lines are the diffusion coefficients which do depend on the
particular system.


\begin{figure}[tbp]
   \begin{center}
      \includegraphics[width=3in]{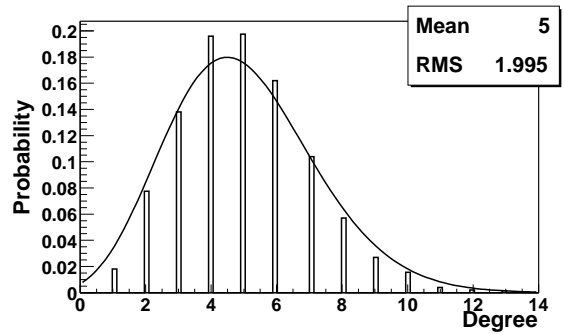}
\caption{A sample degree distribution generated from a random
graph with $5000$ edges and $2000$ vertices as described in the
text.  The line is a Poisson distribution with the same mean value
as the data.}
      \label{fig:PlotDegreeDistribution}
   \end{center}
\end{figure}

Now, consider some graphs which are not regular and, hence, can
produce anomalous diffusion.  One way to generate such graphs is
to begin with a set of vertices.  Next, go through all the
vertices once and add edges where one endpoint of the edge is on
the current vertex and the other goes to a random vertex. This
guarantees that no vertices are isolated.  Then, randomly connect
pairs of vertices until the desired number of edges are laid down.
The probability that a random vertex has $k$ edges is given by
\begin{equation}
   \label{eq:Binomial}
   p(k-1) = \binom{E}{k} \left( \frac{\mu}{V-1}\right)^{k} \left(
1-\frac{\mu}{V-1}\right)^{E-k},
\end{equation}
where $E$ is the number of edges, and $V$ is the number of
vertices and $\mu=2E/V$ is the average number of edge
endpoints per vertex. As the number of edges increases, $p(k)$ in
equation \ref{eq:Binomial} approaches a Poisson distribution with
mean $\mu$, i.e. $p(k) \rightarrow \left(\mu^{k}/k!\right)e^{-\mu}$.
See Figure \ref{fig:PlotDegreeDistribution}.

\begin{figure}[tbp]
   \begin{center}
      \includegraphics[width=3in]{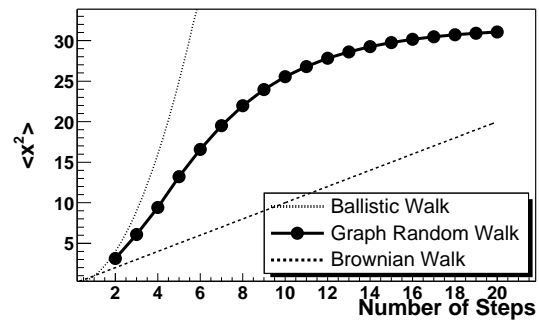}
\caption {A sample plot of $\Bracket{x^{2}(n)}$ for a graph with
degree distribution with a mean of $4$.  $\Bracket{x^{2}(n)}$ for
ballistic walks and Brownian walks are included for reference. The
lines are guides to the eye.}
      \label{fig:PlotSingleXSquared}
   \end{center}
\end{figure}

Once the random graph is constructed, apply the path counting
algorithm and calculate $\Bracket{x^{2}(n)}$. As an example, the
result of this calculation for a graph with average degree $4$ is
shown in Figure \ref{fig:PlotSingleXSquared}. For comparison the
values of $\Bracket{x^{2}(n)}$ are included for a ballistic walk,
$\Bracket{x^{2}(n)} \sim n^{2}$, and a Brownian walk
$\Bracket{x^{2}(n)} \sim n$. As can be seen in Figure
\ref{fig:PlotSingleXSquared}, the value of $\Bracket{x^{2}(n)}$ is
somewhere between a Brownian walk and a ballistic walk so it is
anomalous as expected.

\begin{figure}[tbp]
   \begin{center}
      \includegraphics[width=3in]{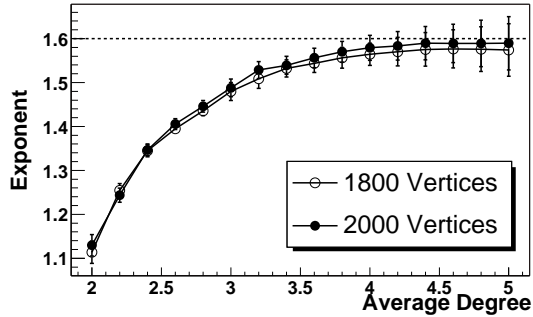}
\caption { A plot of the diffusion exponent versus the average
degree for random graphs with $1800$ vertices and $2000$ vertices.
Each are averaged over three separate realizations. Note that the
anomalous exponent can be controlled by the average degree.
Errors are from fitting $\Bracket{x^{2}(n)}$ to find the exponent.
The lines are guides to the eye.}
      \label{fig:ExponentVsEdge}
   \end{center}
\end{figure}

For a large enough number of steps, $\Bracket{x^{2}(n)}$ of a
random graph saturates because of a finite size effect. For finite
random graphs such as these, there exists a finite average
distance $\overline{D}_{max}$ between any two vertices which
scales like the $\log$ of the number of vertices
\cite{Newman2001b}. This means that once the walker has moved a
distance equal to $\overline{D}_{max}$, every new vertex to which
the walker moves is still only on average a distance
$\overline{D}_{max}$ away from the starting vertex. So
$\Bracket{x^{2}(n)}$ approaches a constant as the number of steps
approaches $\overline{D}_{max}$. Below this plateau
$\Bracket{x^{2}(n)}$ follows a power law $<x^{2}>\sim n^{\alpha}$.
Figure \ref{fig:ExponentVsEdge} shows a plot of the exponent
$\alpha$ from a power law fit for graphs with various average
degrees. As can be seen from the plot, the anomalous diffusion
exponent is controlled by the average degree. Also, as the average
degree increases, the degree distribution approaches a Poisson
distribution. At the same time, the power law for
$\Bracket{x^{2}(n)}$ approaches $1.6$.  Thus, the exponent for a
highly connected random graph with a Poisson degree distribution
is $1.6$.

A problem with modelling random walks with graphs is that the
graph distance does not necessarily correspond to any real spatial
distance. For clarification, imagine an embedding of a graph into
a flat surface, i.e. random dots connected with lines on a piece
of paper. Then, one can calculate $\Bracket{x^{2}(n)}$ directly by
just following a large number of walkers and averaging their
Euclidean distance on the surface measured from where they started
as a function of time. Next, repeat this process for a new
embedding of the same graph with different distances between
the dots but the same connecting lines.  It seems reasonable that this
disorder averaging will lead to a well-defined value for the
exponent. This can then be compared to the results from the
adjacency matrix algorithm using the graph distance. The hope is
that since both give well-defined power laws for the displacement
squared, these two measurements will lead to the same power law,
possibly with a different coefficient but with the same exponent.
To justify this rigorously, one would have to show that the
specific distribution of step sizes in the Euclidean embedding
does not affect the diffusion exponent. In other words, graphs
only keep track of the number of choices a walker has at each
step, not the spatial distance it can go. So for the exponent
calculated with the adjacency matrix algorithm to give the same
results as a random walk in real space, the number of choices must
be the only thing that matters.

One verification that the exponent is insensitive to the details
of the spatial distances comes from our numerical calculation of
the exponent for different realizations of random graphs with a
given degree distribution.  We find that the exponents obtained
from these agree with each other within the error bars of the fit.
On the other hand, the exponent changes when the degree
distribution is changed. This means that it is the degree
distribution, the distribution of the number of choices a walker
has at each step, not the specific realization of the random graph
which determines the exponent.

In conclusion, random walks on random graphs exhibit
characteristics of anomalous diffusion which can be controlled by
the degree distribution of the graph.  Since this is drastically
different from what one finds for a continuum, it seems that
continua and random graphs are fundamentally different structures.
Therefore, when restrictions are placed upon the ability of
diffusing particles to sample their surroundings, anomalous
behavior arises.


\begin{thebibliography}{9}
\expandafter\ifx\csname natexlab\endcsname\relax\def\natexlab#1{#1}\fi
\expandafter\ifx\csname bibnamefont\endcsname\relax
  \def\bibnamefont#1{#1}\fi
\expandafter\ifx\csname bibfnamefont\endcsname\relax
  \def\bibfnamefont#1{#1}\fi
\expandafter\ifx\csname citenamefont\endcsname\relax
  \def\citenamefont#1{#1}\fi
\expandafter\ifx\csname url\endcsname\relax
  \def\url#1{\texttt{#1}}\fi
\expandafter\ifx\csname urlprefix\endcsname\relax\def\urlprefix{URL }\fi
\providecommand{\bibinfo}[2]{#2}
\providecommand{\eprint}[2][]{\url{#2}}

\bibitem[{\citenamefont{Moore and Newman}(2000)}]{Newman2000}
\bibinfo{author}{\bibfnamefont{C.}~\bibnamefont{Moore}} \bibnamefont{and}
  \bibinfo{author}{\bibfnamefont{M.~E.~J.} \bibnamefont{Newman}},
  \bibinfo{journal}{Physical Review E} \textbf{\bibinfo{volume}{62}},
  \bibinfo{pages}{7059} (\bibinfo{year}{2000}).

\bibitem[{\citenamefont{Liljeros et~al.}(1998)\citenamefont{Liljeros, Edling,
  Amaral, Stanley, and $\AA$berg}}]{Liljeros1998}
\bibinfo{author}{\bibfnamefont{F.}~\bibnamefont{Liljeros}},
  \bibinfo{author}{\bibfnamefont{C.~R.} \bibnamefont{Edling}},
  \bibinfo{author}{\bibfnamefont{L.~A.~N.} \bibnamefont{Amaral}},
  \bibinfo{author}{\bibfnamefont{H.~E.} \bibnamefont{Stanley}},
  \bibnamefont{and}
  \bibinfo{author}{\bibfnamefont{Y.}~\bibnamefont{$\AA$berg}},
  \bibinfo{journal}{Science} \textbf{\bibinfo{volume}{280}},
  \bibinfo{pages}{98} (\bibinfo{year}{1998}).

\bibitem[{\citenamefont{Lawrence and Giles}(1998)}]{Lawrence1998}
\bibinfo{author}{\bibfnamefont{S.}~\bibnamefont{Lawrence}} \bibnamefont{and}
  \bibinfo{author}{\bibfnamefont{C.~L.} \bibnamefont{Giles}},
  \bibinfo{journal}{Science} \textbf{\bibinfo{volume}{280}},
  \bibinfo{pages}{98} (\bibinfo{year}{1998}).

\bibitem[{\citenamefont{Albert et~al.}(2000)\citenamefont{Albert, Jeong, and
  Barab$\acute{a}$si}}]{Albert2000}
\bibinfo{author}{\bibfnamefont{R.}~\bibnamefont{Albert}},
  \bibinfo{author}{\bibfnamefont{H.}~\bibnamefont{Jeong}}, \bibnamefont{and}
  \bibinfo{author}{\bibfnamefont{A.-L.} \bibnamefont{Barab$\acute{a}$si}},
  \bibinfo{journal}{Nature} \textbf{\bibinfo{volume}{406}},
  \bibinfo{pages}{378} (\bibinfo{year}{2000}).

\bibitem[{\citenamefont{Albert and Barab$\acute{a}$si}(2002)}]{Albert2002}
\bibinfo{author}{\bibfnamefont{R.}~\bibnamefont{Albert}} \bibnamefont{and}
  \bibinfo{author}{\bibfnamefont{A.-L.} \bibnamefont{Barab$\acute{a}$si}},
  \bibinfo{journal}{Review of Modern Physics} \textbf{\bibinfo{volume}{74}},
  \bibinfo{pages}{47} (\bibinfo{year}{2002}).

\bibitem[{\citenamefont{Sokolov et~al.}(1997)\citenamefont{Sokolov, Mai, and
  Blumen}}]{Sokolov1997}
\bibinfo{author}{\bibfnamefont{I.~M.} \bibnamefont{Sokolov}},
  \bibinfo{author}{\bibfnamefont{J.}~\bibnamefont{Mai}}, \bibnamefont{and}
  \bibinfo{author}{\bibfnamefont{A.}~\bibnamefont{Blumen}},
  \bibinfo{journal}{Physical Review Letters} \textbf{\bibinfo{volume}{79}},
  \bibinfo{pages}{857} (\bibinfo{year}{1997}).

\bibitem[{\citenamefont{Bray and Rodgers}(1988)\citenamefont{Bray, and
  Rodgers}}]{Bray1988}
\bibinfo{author}{\bibfnamefont{A.~J.} \bibnamefont{Bray}},
  \bibnamefont{and}
  \bibinfo{author}{\bibfnamefont{G.~J.}~\bibnamefont{Rodgers}},
  \bibinfo{journal}{Physical Review B} \textbf{\bibinfo{volume}{38}},
  \bibinfo{pages}{11461} (\bibinfo{year}{1988}).

\bibitem[{\citenamefont{Godsil and Royle}(2001)}]{Godsil2001}
\bibinfo{author}{\bibfnamefont{C.}~\bibnamefont{Godsil}} \bibnamefont{and}
  \bibinfo{author}{\bibfnamefont{G.}~\bibnamefont{Royle}},
  \emph{\bibinfo{title}{Algebraic Graph Theory}}, Graduate Texts in Mathematics
  (\bibinfo{publisher}{Springer}, \bibinfo{year}{2001}).

\bibitem[{\citenamefont{Reif}(1965)}]{Reif1965}
\bibinfo{author}{\bibfnamefont{F.}~\bibnamefont{Reif}},
  \emph{\bibinfo{title}{Fundamentals of Statistical and Thermal Physics}}
  (\bibinfo{publisher}{McGraw-Hill Inc.}, \bibinfo{year}{1965}).

\bibitem[{\citenamefont{Newman et~al.}(2001)\citenamefont{Newman, Strogatz, and
  Watts}}]{Newman2001b}
\bibinfo{author}{\bibfnamefont{M.~E.~J.} \bibnamefont{Newman}},
  \bibinfo{author}{\bibfnamefont{S.~H.} \bibnamefont{Strogatz}},
  \bibnamefont{and} \bibinfo{author}{\bibfnamefont{D.~J.} \bibnamefont{Watts}},
  \bibinfo{journal}{Physical Review E} \textbf{\bibinfo{volume}{64}},
  \bibinfo{pages}{026118} (\bibinfo{year}{2001}).

\end{thebibliography}

\end{document}